\documentclass[twocolumn]{webofc}

\usepackage[varg]{txfonts}   
\usepackage{hyperref}
\usepackage{url}
\usepackage{lineno, blindtext, indentfirst}
\usepackage{graphicx}
\usepackage{floatrow}
\usepackage{multirow}
\hypersetup{colorlinks=true,citecolor=blue,urlcolor=blue,linkcolor=blue}

\newcommand{ \pXi }{$p$$-$$\Xi^{-}$}
\newcommand{ \LaLa }{$\Lambda$$-$$\Lambda$}
\newcommand{ \pOm }{$p$$-$$\Omega^{-}$}
\newcommand{ \pLa }{$p$$-$$\Lambda$}
\begin{document}
    
%
\title{Femtoscopy of Strange Baryons in Heavy-ion Collisions at RHIC-STAR}
%
%

\author{\firstname{Boyang} \lastname{Fu for the STAR Collaboration}\inst{1}\fnsep\thanks{\email{boyangfu@mails.ccnu.edu.cn}} 
}

\institute{Institute of Particle Physics and Key Laboratory of Quark \& Lepton Physics (MOE), \\Central China Normal University, Wuhan, 430079, China. 
\
          }

\abstract{ Studying the final state interactions and finding possible bound states is helpful for understanding the strong interactions and comprehending the equation-of-state (EoS) of the nuclear matter. In these proceedings, we present recent femtoscopy results of \pXi{}, \LaLa{}, \pOm{} femtoscopic correlations with high statistics Isobar (Ru+Ru, Zr+Zr) and Au+Au collisions measured by the STAR experiment. For the \pXi{} and \pOm{} pairs, the centrality dependence of source size and the scattering parameters are extracted with the Lednický-Lyuboshitz approach. The results show that there is an attractive interaction in \pXi{} pairs and a bound state in \pOm{} pairs.
}
\maketitle
\section{Introduction}
\label{intro}
\setlength{\parindent}{0em}
Baryon interactions are of fundamental interest in nuclear physics \cite{14,15,16} and astrophysics. Further, understanding the Equation of State (EoS) of the nuclear matter \cite{1,2} at high baryon density is important to understand the properties of neutron stars to solve the hyperon puzzle. Seraching for dibaryon bound states is also of fundamemtal interest. The H-dibaryon—a deeply bound six-quark state—was first proposed by Jaffe in 1977 \cite{3}. Another candidate, the nucleon–$\Omega\ (\text{N}\Omega)$ dibaryon, has also attracted significant interest \cite{4}.

\label{sec-2}
In heavy-ion collisions, two-particle femtoscopy is a powerful and unique method for extracting information about the spatio-temporal properties of the source, the final state interactions among the particles considered and searching for the possible bound states. In these proceedings, we present new results of \pXi{}, \LaLa{}, \pOm{} correlation in AuAu and Isobar collisions at $\sqrt{s_{NN}}$ = 3 GeV and 200 GeV from RHIC-STAR and use Lednický-Lyuboshitz (LL) formalism \cite{5} to extract scattering lengths ($f_0$) and effective ranges ($d_0$), discussing their implications for possible strange dibaryon bound states with strangeness $S = -2\ $and$\ S = -3$.

\section{Analysis Details}
In this analysis, particle identification was performed using the Time-Projection-Chamber (TPC) and Time-of-Flight (TOF) detectors. For decay daughters, reconstruction was carried out via the helix swimming method. Specifically, $\Xi^{-}$ candidates were reconstructed through the decay channel $\Xi^{-} \rightarrow\ \Lambda\ +\ \pi^{-}$ , $\Lambda$ candidates via $\Lambda \rightarrow\ p\ +\ \pi^{-}$, and $\Omega^{-}$ candidates through the decay channel $\Omega^{-} \rightarrow\ \Lambda\ +\ K^{-}$.

\begin{figure}[H]
    \centering
        \includegraphics[width=\columnwidth]{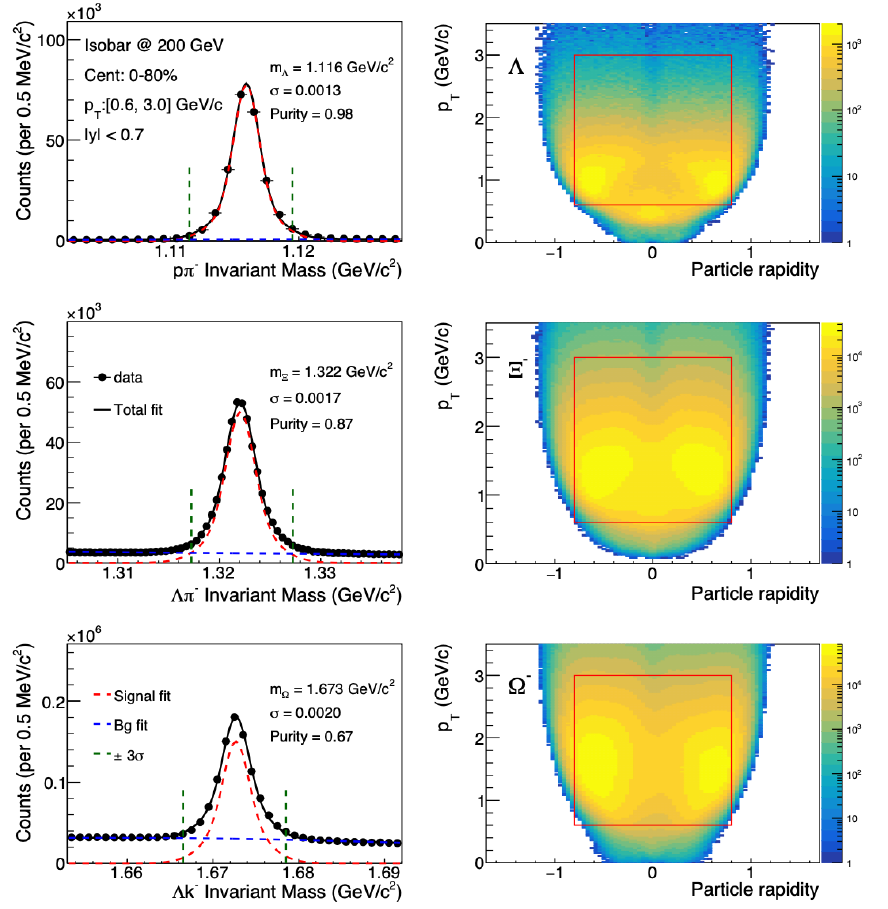}
        \caption{The left three figures are the examples of purity calculation of different particles with multi-gaussian function and polynomial function. The right three figures are the acceptance of three particles respectively.}
        \label{fig:pid}
\end{figure}

The observable of interest in femtoscopy is the two-particle correlation function. The correlation function can be expressed theoretically as $C(k^{*})=\int d^{3}r^{*}S(r^{*})\lvert \Psi(r^{*},k^{*})\rvert^{2}$, where $k^{*}$ and $r^{*}$ are the relative momentum and relative distance of the pair of interest in the Pair rest Frame (PRF). $S(r^{*})$ is the source function and $\Psi(r^{*},k^{*})$ represents the wave function of the pair of interest. Experimentally, this correlation function is computed as $C(k^{*})=\mathcal{N}[A(k^{*})/B(k^{*})]$, where $A(k^{*})$ is the correlated pairs in same-event, and $B(k^{*})$ is the un-correlated background pairs obtained from mixed-events. The normalization parameter $\mathcal{N}$ is chosen such that the mean value of the correlation function equals unity for large $k^{*}$ range.

The raw correlation functions are corrected to get the genuine correlation function. Detector effects, which include track merging and track splitting effect, are removed with $\Delta\Phi^{*} vs\ \Delta\eta$ cuts. Background correlation functions are estimated by side-band method. Purity and feed-down corrections are both considered. In this analysis, residual correlation is also removed.

\section{Results}
The correlation function is parameterized using the Lednický-Lyuboshitz (LL) model, which considers a static spherical Gaussian source under a smoothness approximation, convoluted with an S-wave function. In LL model, the complex scattering amplitude $f^{S}(k^*)$ with Coulomb interaction is evaluated via the effective range approximation,
\begin{equation}
   f^{S}(k^{*})=\left[\frac{1}{f_{0}^{S}}+\frac{1}{2}d_{0}^{S}k^{*2}-\frac{2}{a_{c}}h(\eta)-ik^{*}A_{c}(\eta)\right]^{-1}
\end{equation}
where $f_{0}^{S}$ is scattering length, $d_{0}^{S}$ is effective range of the interaction, $a_{c}$ is the Bohr radius, $h(\eta)$ is complex function, $\eta$ and $A_{c}$ are Coulomb related factor, and $S$ denotes the total spin of the pair.

\subsection{\pXi{} Correlation Function}

\begin{figure}[H]
    \centering
        \includegraphics[width=\columnwidth]{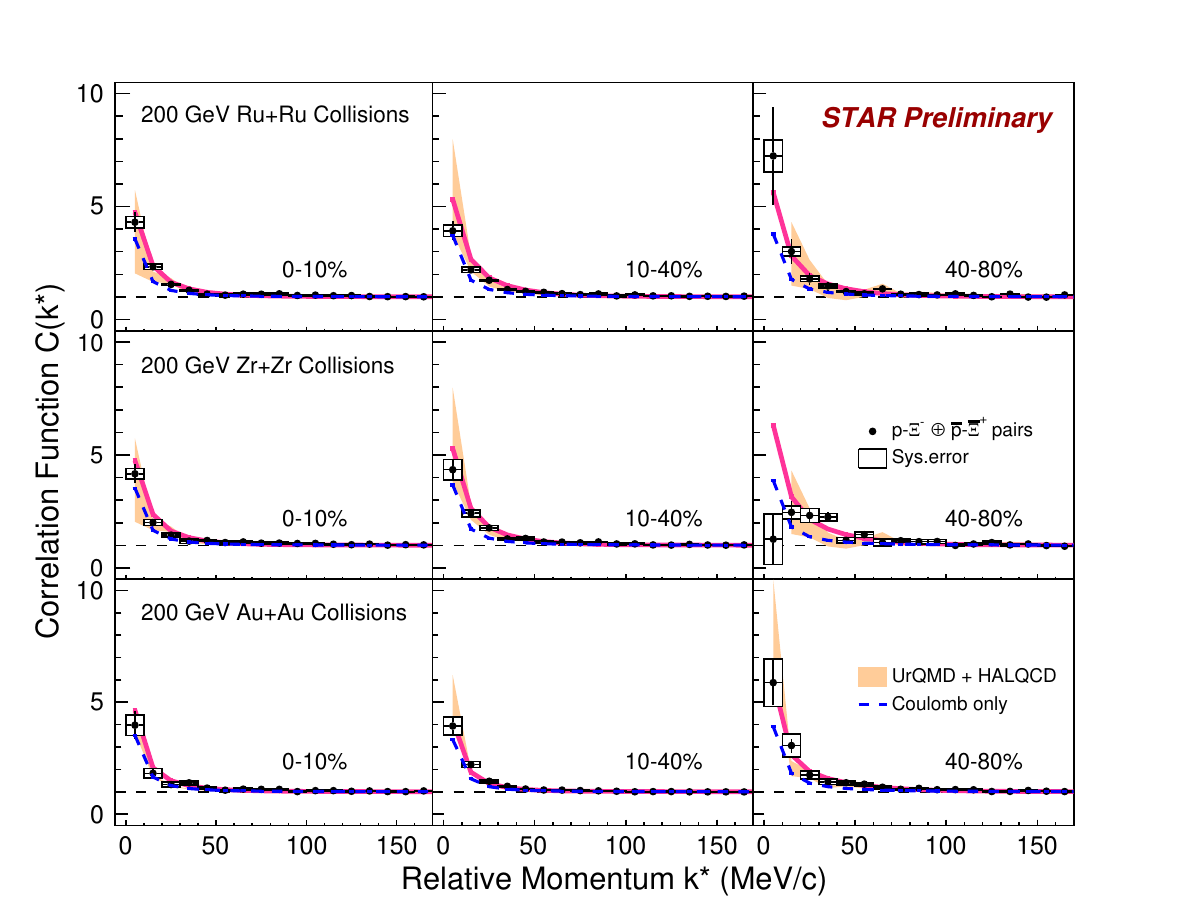}
        \caption{Measured \pXi{} correlation functions in 0-10\% (left), 10-40\% (middle) and 40-80\% (right) centrality in Ru+Ru (top), Zr+Zr (middle) and Au+Au (bottom) collisions at $\sqrt{s_{\mathrm{NN}}}$ = 200 GeV. Black bars and boxes represent the statistical and systematic uncertainties. Magenta lines show simultaneous fits with Lednický-Lyuboshitz model (including effects of Coulomb and strong interactions), blue dashed lines show pure Coulomb and orange bands represent the results from UrQMD + HAL QCD simulation.}
        \label{fig:pXi_CF}
\end{figure}

\begin{table}[H]
\centering
\begin{tabular}{|c|c|c|c|}
\hline
\text{LL Fit} & \text{$f_{0}$ (fm)} & \text{$d_{0}$ (fm)} & \text{$\chi^{2}$/ndf} \\
\hline
Ru+Ru & \multirow{3}{*}{$0.69^{+0.11}_{-0.10}$} & \multirow{3}{*}{$12.60^{+5.12}_{-7.00}$} & \multirow{3}{*}{1.23} \\
\cline{1-1}
Zr+Zr & & & \\
\cline{1-1}
Au+Au & & & \\
\hline
\end{tabular}
\caption{Extracted scattering length and effective range of \pXi{} pairs from LL model fitting.}
\label{tab:pxi}
\end{table}

\begin{figure}[htbp]
    \centering
        \includegraphics[width=\columnwidth]{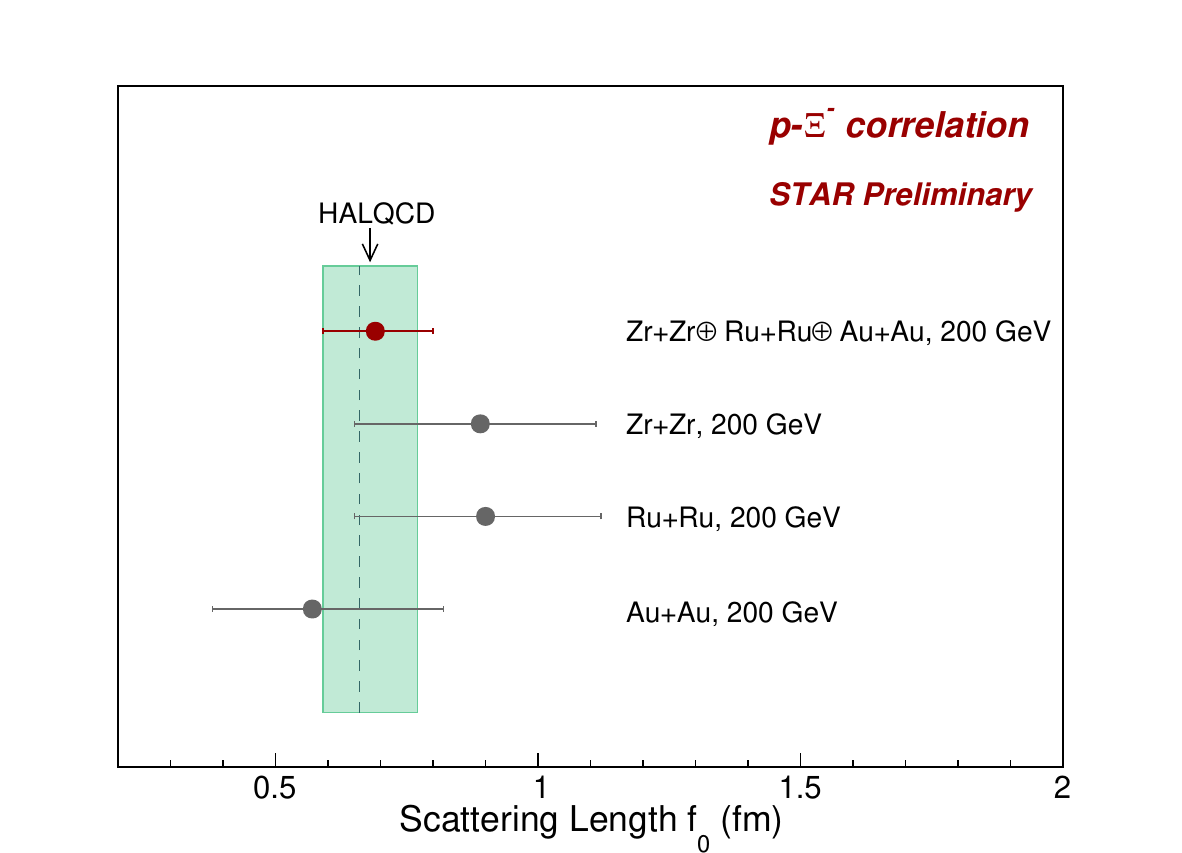}
        \caption{The \pXi{} spin-averaged $f_0$ obtained from simultaneously LL fit in Isobar and Au+Au collisions by Bayesian method. Red solid points represent the results from simultaneous fits. The black points indicate the results for the three individual systems. The green band shows the prediction from HAL QCD.}
        \label{fig:pXi_f0}
\end{figure}

\setlength{\parindent}{0em}
Fig.~\ref{fig:pXi_CF} shows the measured \pXi{} correlation function as a function of $k^*$ in three centrality bins in Isobar and Au+Au collisions at $\sqrt{s_{\mathrm{NN}}}$ = 200 GeV. The measured correlation functions show clear enhancement at low $k^*$ in all centrality classes and become more pronounced in peripheral collisions. It is clear that only Coulomb interaction cannot account for the full measured correlation functions. Results of the UrQMD + HAL QCD calculations \cite{7,8} and LL fitting results both describe the data reasonably well in all centrality classes. Fig.~\ref{fig:pXi_f0} shows the scattering length ($f_0$) extracted from LL simultaneous fits to data in different systems and centrality classes with a common set of parameters ($f_0$, $d_0$) and different emission source sizes ($R_G$). The positive $f_0$ supports a weakly attractive interaction in \pXi{} pairs. The results are summarized in Tab.~\ref{tab:pxi}.

\subsection{\LaLa{} Correlation Function}
\setlength{\parindent}{0em}

\begin{figure}[htbp]
    \centering
        \includegraphics[width=0.85\columnwidth]{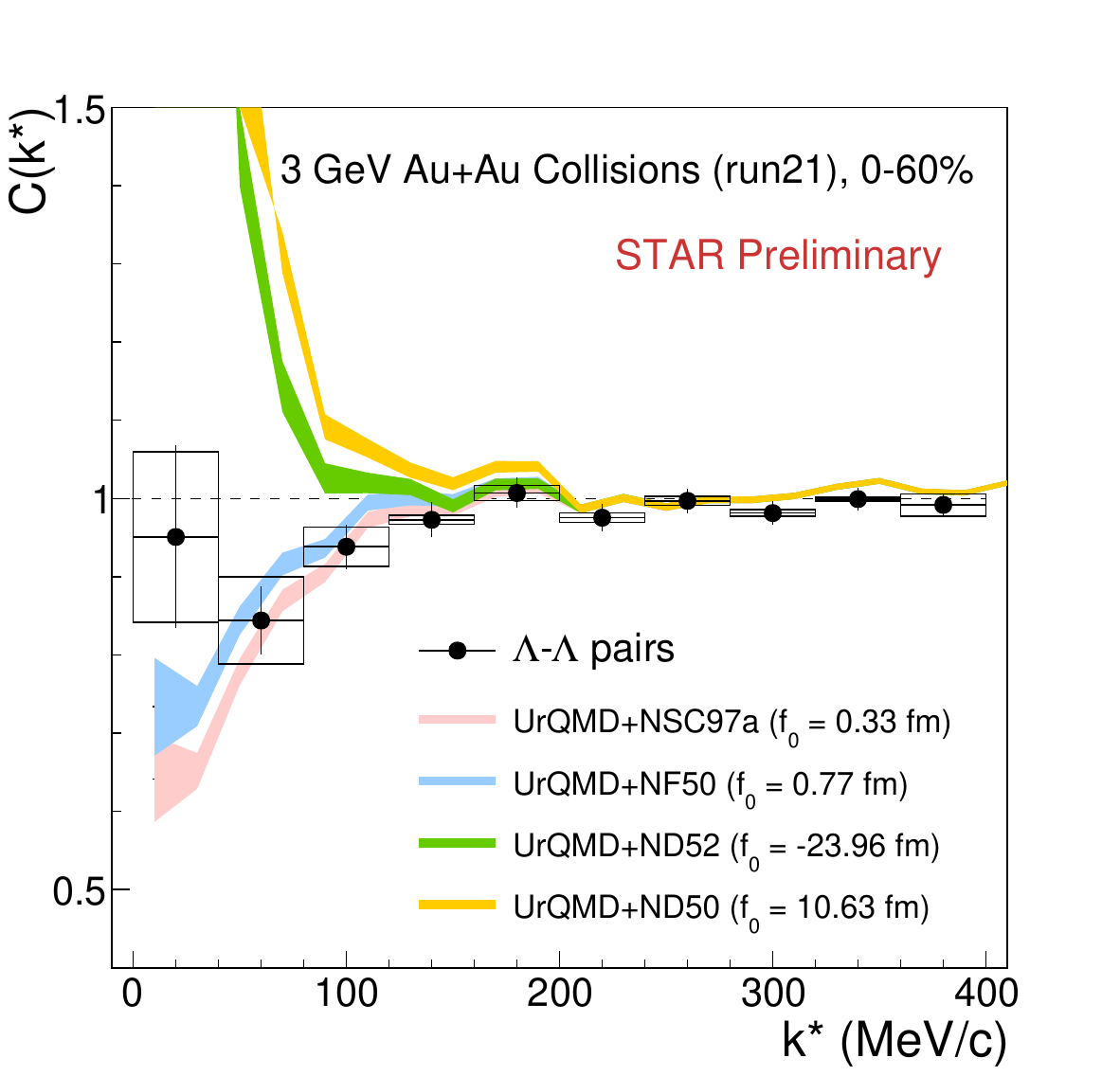}
        \caption{Measured \LaLa{} correlation functions in 0-60\% centrality in Au+Au collisions at $\sqrt{s_{\mathrm{NN}}}$ = 3 GeV. Black bars and boxes represent the statistical and systematic uncertainties. Different colored lines represent UrQMD model simulation with different potentials given by different theories.}
        \label{fig:LaLa_CF}
\end{figure}

\begin{table}[H]
\centering
\begin{tabular}{|c|c|c|c|}
\hline
\text{Potential} & \text{$f_{0}$ (fm)} & \text{$d_{0}$ (fm)} & \text{$\chi^{2}$/ndf} \\
\hline
NSC97a & 0.33 & 12.37 & 1.53 \\
\hline
NF50 & 0.77 & 4.27 & 1.61 \\
\hline
ND52 & -23.96 & 2.59 & 2.24 \\
\hline
ND50 & 10.63 & 2.04 & 4.02 \\
\hline
\end{tabular}
\caption{Interaction parameters pf \LaLa{} from different potential models.}
\label{tab:lala}
\end{table}

Fig.~\ref{fig:LaLa_CF} shows the measured \LaLa{} correlation function as a function of $k^*$ in 0-60\% centrality bin in Au+Au collisions at $\sqrt{s_{\mathrm{NN}}}$ = 3 GeV. The correlation function shows suppression at low $k^*$. The correlation function is compared with UrQMD model with different potentials \cite{9,10,11} which are shown as different colored lines. The parameters of different potentials are summarized in Tab.~\ref{tab:lala} The result shows the simulation with positive $f_0$ is in better agreement with data which hints at an attractive interaction in \LaLa{} pairs.

\subsection{\pOm{} Correlation Function}

\begin{figure}[H]
    \centering
    \includegraphics[width=\columnwidth]{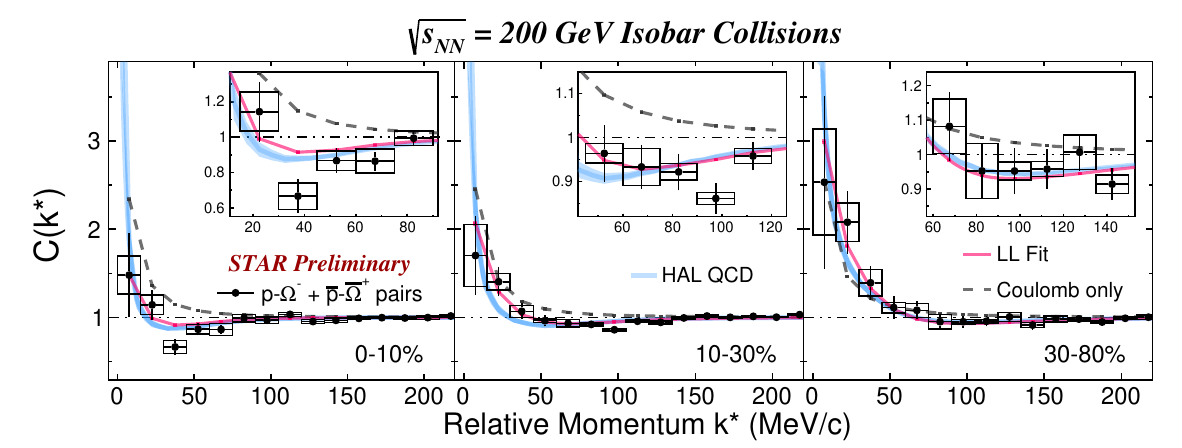}
    \caption{Correlation functions for \pOm{} measured in Ru+Ru and Zr+Zr collisions at $\sqrt{s_{\mathrm{NN}}}$ = 200 GeV. The magenta lines show fits using the LL model with the spin-averaged method, while gray dashed lines represent Coulomb-only contributions. Blue bands indicate HAL QCD predictions. Insets zoom in near unity.}
    \label{fig:pOm_CF}
\end{figure}

Fig.~\ref{fig:pOm_CF} shows the \pOm{} correlation functions in Ru+Ru and Zr+Zr collisions at $\sqrt{s_{\mathrm{NN}}}$ = 200 GeV in three different centrality intervals. Similar to \pXi{}, an enhancement appears at low $k^*$ together with a suppression below unity (see insets). This suppression suggests strong interaction effects from either a repulsive core or a bound-state. 

\begin{figure}[H]
    \centering
    \includegraphics[width=\columnwidth]{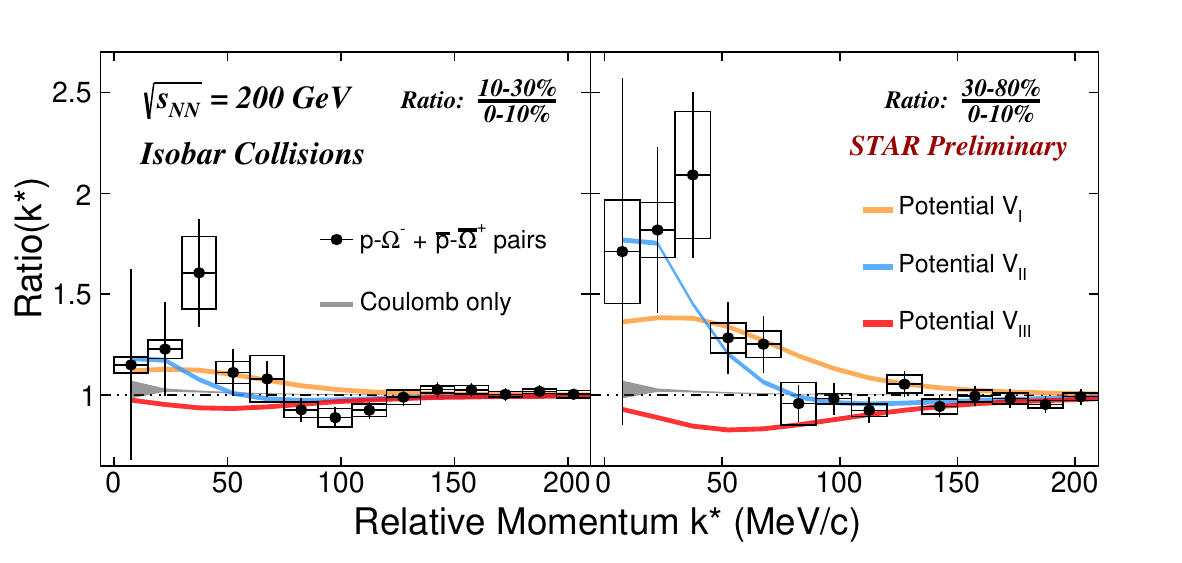}
    \caption{The ratio of correlation functions in different centralities. Different colorful lines represent different model calculations using different potentials.}
    \label{fig:pOm_ratio}
\end{figure}

\begin{table}[H]
\centering
\begin{tabular}{|c|c|c|c|c|}
\hline
\text{Potential} & \text{$f_{0}$ (fm)} & \text{$d_{0}$ (fm)} & BE (MeV) & \text{$\chi^{2}$/ndf} \\
\hline
$V_{I}$ & 1.12 & 1.16 & -- & 1.66 \\
\hline
$V_{II}$ & -3.38 & 1.13 & 2.15 & 0.76 \\
\hline
$V_{III}$ & -1.29 & 0.65 & 26.9 & 2.02 \\
\hline
\end{tabular}
\caption{Interaction parameters of \pOm{} from different potential models.}
\label{tab:ratio}
\end{table}

Fig.~\ref{fig:pOm_ratio} shows the ratio of correlation functions of $\frac{10-30\%}{0-10\%}$ and $\frac{30-80\%}{10-30\%}$. By taking correlation function ratio, coulomb effect can be largely canceled. The ratio shows enhancement at low k* range and depletion around k*~100 MeV/$c$ which is due to the presence of shallow bound state. The potentials information is summarized in Tab.~\ref{tab:ratio} The potential $V_{II}$ provides a better description of the data.

\begin{figure}[H]
    \centering
    \includegraphics[width=\columnwidth]{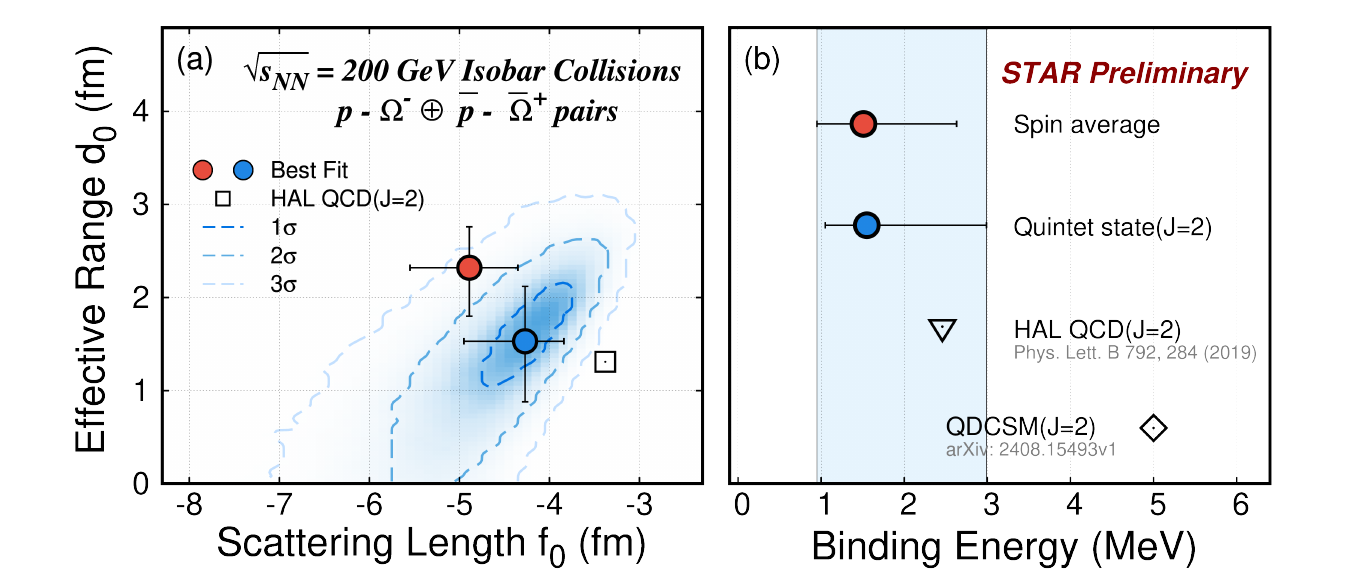}
    \caption{(a)The extracted \pOm{} scattering parameters $f_0$ and $d_0$, are shown as probability contours from spin-averaged (red) and quintet (blue) method. Blue bands show 1–3$\sigma$ confidence levels from the quintet method. (b)The \pOm{} binding energy calculated from the extracted $f_0$ and $d_0$ is shown for the spin-averaged and quintet fits.}
    \label{fig:pOm_f0}
\end{figure}

Fig.~\ref{fig:pOm_f0} presents the extracted $f_0$ and $d_0$ from both spin-averaged and quintet-channel fits—the latter treating quintet states with strong plus Coulomb interactions and triplet states with only Coulomb interactions. The right panel shows the binding energy (E) calculated from extracted $f_0$ and $d_0$ using the Bethe formula \cite{17}, indicating a shallow bound state consistent with HAL QCD calculations \cite{12,13}. The results are summarized in Tab.~\ref{tab:pOm}.

\clearpage
\noindent
\begin{minipage}[t][0.48\textheight]{0.48\textwidth}
\vspace{0pt}  

\begin{table}[H]
\centering
\begin{tabular}{|c|c|c|c|c|}
\hline
\text{} & \text{Spin ave.} & \text{Quintet} & \text{HAL QCD} \\
\hline
$f_{0}$ (fm) & $-4.9^{+0.5}_{-0.7}$ & $-4.3^{+0.4}_{-0.7}$ & -3.4 \\
\hline
$d_{0}$ (fm) & $2.3^{+0.4}_{-0.5}$ & $1.5^{+0.5}_{-0.7}$ & 1.3 \\
\hline
BE (MeV) & $1.5^{+1.1}_{-0.6}$ & $1.6^{+1.4}_{-0.5}$ & 2.3 \\
\hline
\end{tabular}
\caption{Extracted scattering length and effective range of \pOm{} pairs from LL model fitting.}
\label{tab:pOm}
\end{table}

\subsection{Extracted Source and Scattering Parameters}

Fig.~\ref{fig:f0d0_all} shows the extracted scattering parameters for the \pLa{}, \pXi{}, \pOm{} pairs. Fig.~\ref{fig:Source} presents the extracted source size as a function of charged multiplicity density $(\frac{dN_{ch}}{d\eta})^{1/3}$. The extracted source sizes fall within a reasonable range and show a clear centrality dependence, with more central collisions corresponding to larger source sizes.Fig.~\ref{fig:f0d0_all} shows the extracted scattering parameters for the \pLa{}, \pXi{}, \pOm{} pairs. 
\begin{figure}[H]
    \centering
        \includegraphics[width=\columnwidth]{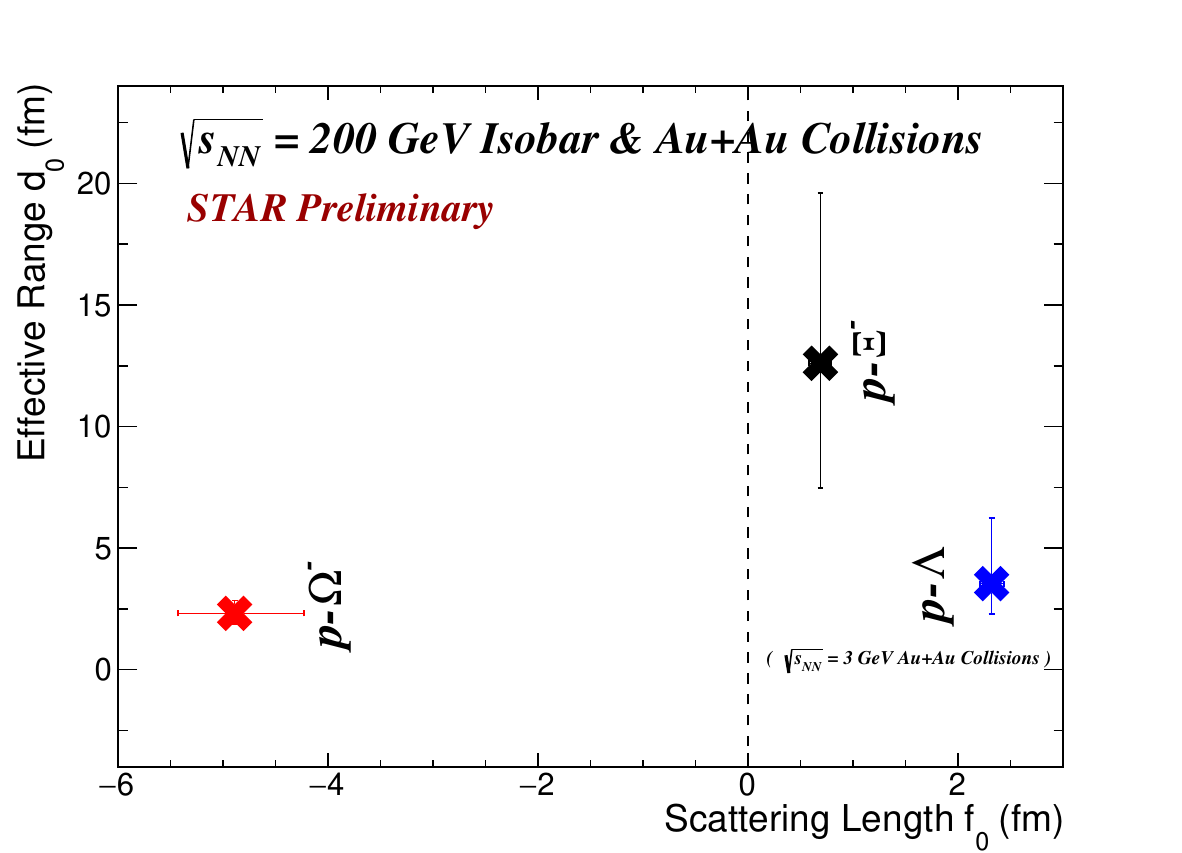}
        \caption{Extracted final state interaction parameters: $f_0$ and $d_0$ for \pLa{} (blue), \pXi{} (black) and \pOm{} (red) pairs.}
        \label{fig:f0d0_all}
\end{figure}

\begin{figure}[H]
    \centering
        \includegraphics[width=\columnwidth]{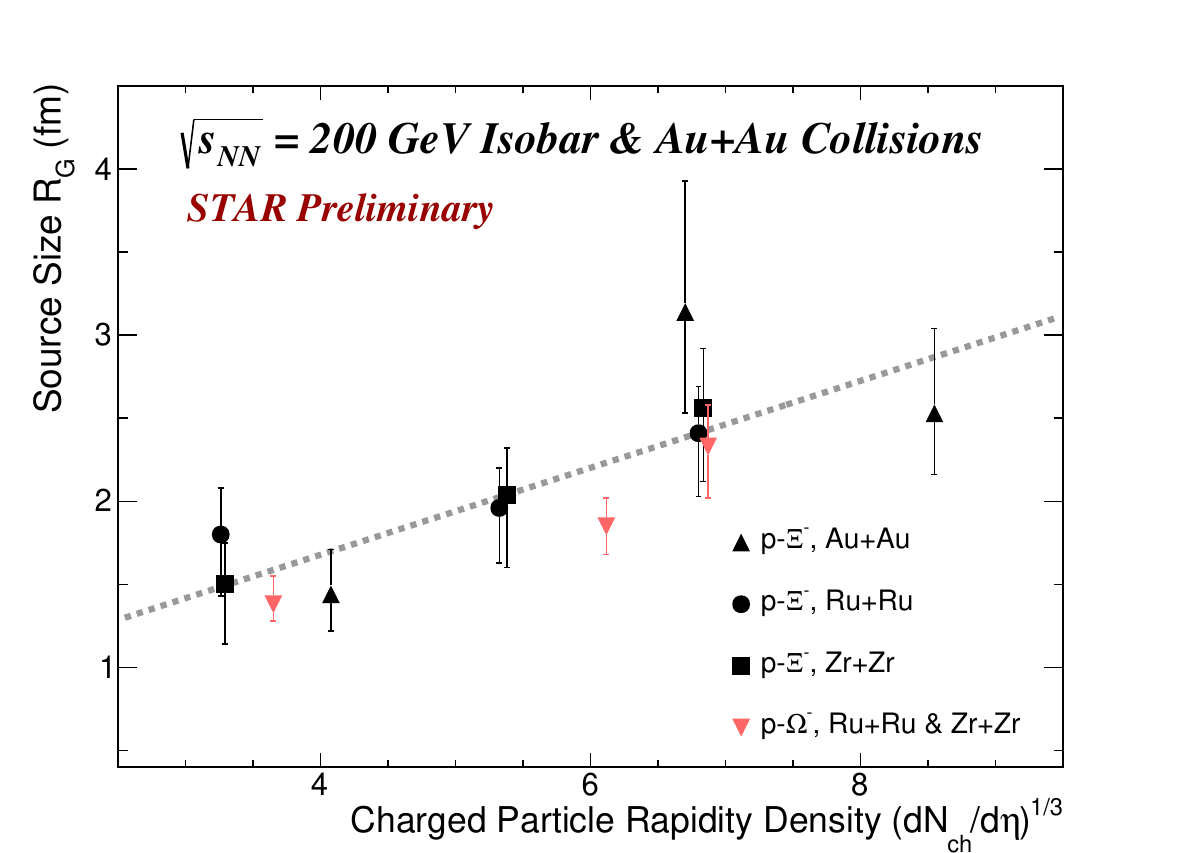}
        \caption{Extracted source size $R_G$ parameter as a function of $(\frac{dN_{ch}}{d\eta})^{1/3}$ for \pXi{} and \pOm{} pairs in different collision systems.}
        \label{fig:Source}
\end{figure}
\end{minipage}

\begin{minipage}[t][0.48\textheight]{0.48\textwidth}
Fig.~\ref{fig:f0d0_all} shows the extracted scattering parameters for the \pLa{}, \pXi{}, \pOm{} pairs. Fig.~\ref{fig:Source} presents the extracted source size as a function of charged multiplicity density $(\frac{dN_{ch}}{d\eta})^{1/3}$. The extracted source sizes fall within a reasonable range and show a clear centrality dependence, with more central collisions corresponding to larger source sizes.

\section{Summary}
\label{summary}
\setlength{\parindent}{0em}

In these proceedings, we present the measurements of \pXi{}, \LaLa{} and \pOm{} correlation functions in Isobar and Au+Au collisions at $\sqrt{s_{\mathrm{NN}}}$ = 200 GeV and 3 GeV. For the first time, the source size and strong interaction parameters are extracted for \pXi{} and \pOm{} pairs in each collision systems with Lednický-Lyuboshitz approach. It is observed that, there is an attractive interaction in \pXi{} pair and probably in \LaLa{} pair. For \pOm{}, a significant suppression at low $k^*$ provides the first experimental evidence for a shallow bound state. 

\section{Acknowledgement}
Thanks to STAR Collaboration. This work was supported in part by National Key Research and Development Program of China (No.2022YFA1604900), National Natural Science Foundation of China (No.12122505 and No.12525509 ).

\end{minipage}

\end{document}